\documentstyle[11pt,newpasp,twoside, epsf]{article}
\begin{document}
\title{Quasars and Narrow-Line Seyfert 1s: Trends and Selection Effects}
\author{Joseph C. Shields and Anca Constantin}
\affil{Physics \& Astronomy Department, Ohio University, Athens, OH 45701}

\begin{abstract}
The SDSS has opened a new era for the study of AGN spectroscopic
properties and how these depend on luminosity and time.  In this
presentation we review some of the current issues and problems in
studies of high-redshift quasars and Narrow-Line Seyfert 1
galaxies. Investigations employing SDSS will, in some situations,
still have to pay attention to selection biases and other fundamental
limitations of the data.
\end{abstract}

\section{Introduction}
The existing literature and pre-Sloan Digital Sky Survey (SDSS) data
pertaining to the emission line properties of AGNs have numerous limitations
in terms of sample selection criteria and homogeneity.
These issues are particularly evident in studies of high redshift quasars,
where the focus is on $z > 4$. In this contribution we discuss some
of the outstanding questions in this area and how these connect to our
understanding of other AGNs, including Narrow-Line Seyfert 1 (NLS1) galaxies.
SDSS will lead to significant improvements in several ways relevant to
these topics, but it is important to recognize some of the concerns and
potential problems that we will still encounter in using SDSS in studies
of this type.

\section{QSO Selection Effects}
During the past decade the number of QSOs known to exist at $z>4$ has
increased rapidly. Much of the early interest in these objects was
directed at simply finding them, and using them as background light
sources for studies of intervening absorbers.  When the number of
known sources grew from a handful to several dozen, Shields and Fred
Hamann initiated a program of optical spectroscopy with the MMT and
Keck in order to study their emission-line properties in more detail;
the results were published in Constantin et al. (2002).  Our data
spanned $\sim 1100 - 1700$\AA\ in the rest frame.

One of the results that came out of our survey was that a significant
fraction of the $z>4$ sources exhibit strong emission lines, with somewhat
greater frequency than is seen in typical samples at $z\sim 2-3$.
This finding is of interest since it could be indicative of evolution
with redshift in quasar properties.  A related point is that linewidth
is inversely correlated with line strength, in this survey as well as
others at lower redshift; this means that the strong-lined objects in
our sample also tend to display relatively narrow linewidths; do the
high-$z$ AGNs have small black hole masses for their luminosity? Mathur
(2000) suggested that this is in fact the case, based on our preliminary
findings.  Her idea was that the high-$z$ sources may show a preference
for NLS1 behavior, and that both types of object may be in an early
evolutionary phase characterized by a high Eddington ratio $L/L_{Edd}$.
The implications for the growth of black holes and the lifecycle of AGNs
are potentially quite significant, if this scenario is true.

The existing samples of $z>4$ QSOs are subject to some difficulties
that bear on this matter, however. The majority of these sources were
discovered by color selection techniques.  The strongest signature of
these objects is a very red color in $B-R$ or equivalent bandpasses,
that results from the Ly$\alpha$ line in $R$ and an attenuated
continuum in $B$ due to the Ly$\alpha$ forest.  The influence of the
Ly$\alpha$ line is heightened at these redshifts by the $1+z$ scaling
of equivalent width.  The result is a potential bias favoring
detection of objects with strong lines: a stronger line produces a
brighter $R$ magnitude for a given AGN continuum level.  In Constantin
et al. we reviewed the issue of selection effects for our $z>4$ sample
but were unable to draw strong conclusions. The extent of bias is
complicated in various ways by details of the process by which QSO
candidates were originally identified, which is not homogeneous and
in some cases not well documented.

A bias toward strong lines will translate into other biases in the
emission-line results for an AGN sample.  Principal Components/Eigenvector
analyses imply that many aspects of quasar spectra are correlated,
including as noted above a connection between line strength and line
width. Selection effects can thus potentially skew survey results
in terms of emission-line ratios, inferred black hole masses, Eddington
ratios, and other properties, if not taken into consideration.

The SDSS offers several obvious improvements to the study of high-$z$
quasars, but will still have some important limitations that should
not be ignored. On the plus side, the SDSS offers multiple colors for
categorizing objects, excellent photometry, and a procedure for
identification of AGN candidates that is well defined and documented.
The identification of AGN samples with this prescription can be
expected to be highly complete; however, this does not mean that
biases can be ignored. As shown by the simulations provided by
Fan et al. (2001), candidate selection is still affected by strong
lines; while the consequences are likely to be minor over most of the
SDSS magnitude range, the possibility of selection bias becomes much
more worrisome near the flux limit. The results presented at this
conference by Dan Vanden Berk suggesting that SDSS quasars at $z > 4$
have larger equivalent widths at a given luminosity than their lower
$z$ counterparts is the sort of finding that should be checked
carefully in this regard.

\section{Comparison of AGN Samples}
Drawing conclusions about AGN evolution requires comparisons of
sources measured across some range of redshifts, which invariably
raises observational challenges. The first is that in the large
majority of samples (certainly the majority of samples currently in
the literature), redshift and luminosity $L$ are highly correlated,
because of their usual linkage in flux-limited samples.  When
differences in spectral properties are found between low- and high-$z$
sample members, it then becomes difficult to ascribe this with
certainty to evolution, since the distinction could be driven by
luminosity differences; conversely, for those interested in AGN
physics as a function of source luminosity, the possibility of
evolutionary effects is a complicating factor.

The solution, of course, is to construct samples with sufficient
coverage of the $L-z$ plane to make it possible to compare objects at
a common $L$ across a range of $z$, or to compare sources at a single
$z$ that span a range of $L$. One of the more successful recent
attempts in this direction was conducted by Dietrich et al. (2002),
who demonstrated rather convincingly that $L$, and not $z$, is the
fundamental parameter underlying the Baldwin Effect (the negative
correlation between luminosity and line equivalent width in AGNs).
Dietrich et al. used spectra drawn from a diverse collection of
ground- and space-based studies.  Quasars in the SDSS still have
substantial correlation between $L$ and $z$, but as shown by Vanden
Berk at this meeting, the coverage of the $L-z$ plane is nonetheless
sufficiently broad to separate the two variables, confirming the
Dietrich et al. results.  An impressive aspect of the SDSS is the
large number of quasar spectra available, which makes it possible to
construct quite narrow bins in the $L-z$ plane that still contain
statistically robust subsamples.

A second observational challenge is that of available wavelength
coverage as a function of $z$. While the optical spectral coverage of
the SDSS is substantial, the rest-frame span of wavelengths we're
looking at is considerably larger, so that comparison of spectral
properties for samples with disparate redshifts requires some extra
work, in terms of observations and/or analysis techniques.  The
following section presents a case study of this type that
we encountered involving the $z>4$ QSO spectra.

\section{NLS1s at High $z$?}
Given the uncertain role of selection effects in the Constantin et al.
QSOs, we thought it would be worthwhile to look for quantitative tests
of the suggestion that these objects are analogs of NLS1s.  Comparisons
between the $z>4$ spectra and most published spectra of NLS1s are impossible
since the former cover restframe UV wavelengths and the latter mostly
span optical intervals.  The NLS1 classification is based on
the H$\beta$ and [OIII]$\lambda$5007 lines, which are very difficult
to measure in the high-$z$ objects.  We consequently resorted to comparisons
in the UV bandpass, using archival HST spectra for a sample of 22 NLS1s.
Details of this study are presented in Constantin \& Shields. (2003).

The simplest way to compare the two samples spectroscopically is by a
direct comparison of composite spectra.  Such an exercise is shown in
Figure 1.  Differences are present, most obviously in the strengths of
the lines, with larger equivalent widths in the NLS1 composite.
Unfortunately it is unclear whether this difference has any bearing on
the NLS1 nature of the QSOs; the NLS1s are substantially lower in
luminosity than the QSOs, and the difference is thus entirely
consistent with the usual Baldwin Effect.

\begin{figure}
\plotone{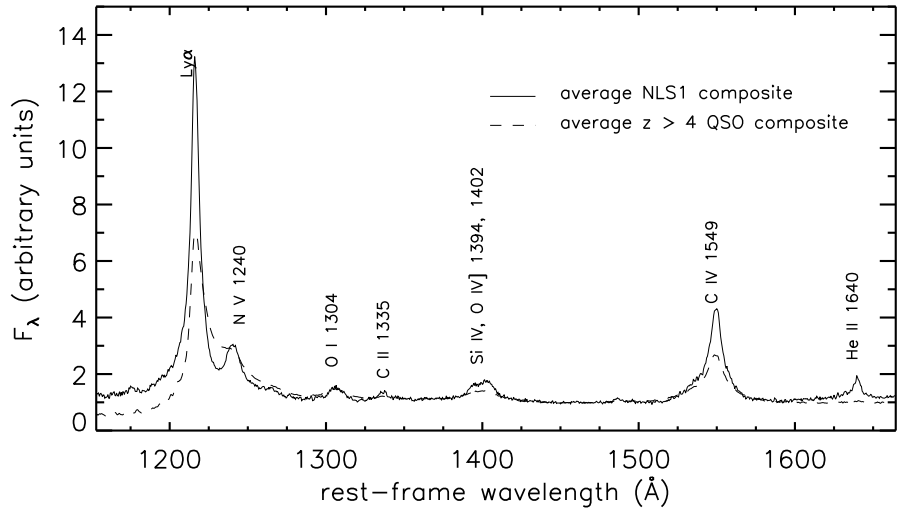}
\caption{Comparison of composite spectra for NLS1s and  $z>4$ QSOs,
from Constantin \& Shields. (2003).}
\end{figure}

Constructing a comparison of the two object types that circumvents
luminosity dependences is difficult.  One possible strategy that we
finally implemented is to use a Principal Component (PC) analysis to
characterize the NLS1 and QSO samples, in order to gauge their similarity.
A diagnostic is then the extent to which a common set of PCs accounts
for the spectra and their total variance.  The conclusion in the end
is that the NLS1s are more spectroscopically ``compact,'' with a larger
fraction of their variance explained with a small number of PCs.
High-$z$ QSOs are probably not close cousins to NLS1s, and whatever
tendency they show towards narrow lines is potentially explainable
as a selection effect, as noted in \S 2.

The SDSS is an incredible resource for identifying new NLS1s as
well as other types of interesting AGN subsamples. As the discussion
above illustrates, however, bandpass limitations for the SDSS data
and luminosity effects will still present challenges in studies of
AGN phenomenology and evolution, requiring supporting UV and IR data
from other sources, as well as creativity in analysis methods.

\question{Antonucci} Anecdotally, high-ionization lines in NLS1s are sometimes
broader and blueshifted relative to low-ionization lines.  Can you tell us
the details and generality of those statements?

\answer{Shields} Our study shows a clear trend that higher ionization
lines exhibit larger blueshifts. This is true both for the broad and narrow
lines.  There is little evidence of a trend in line width.

\end{document}